\documentstyle[12pt]{article}
\begin{document}
\newcommand{\dia}{\mbox{diag} }
\newcommand{\ttr}{\mbox{Tr}\ }
\newcommand{\aaa}{A_\alpha}
\newcommand{\ha}{H_\alpha}
\newcommand{\gab}{g_{\alpha\beta}}
\newcommand{\gba}{g_{\beta\alpha}}
\newcommand{\la}{\Lambda_\alpha}
\newcommand{\be}{\begin{equation}}
\newcommand{\ee}{\end{equation}}
\newcommand{\calh}{{\cal H}}
\newcommand{\ra}{\rho_\alpha}
\newcommand{\ca}{{\cal A}}
\newcommand{\boe}{{\cal B}(E)}
\newcommand{\fuks}{\phi (u,\,x)}
\newcommand{\fit}{\phi (t,\,x)}
\newcommand{\luks}{\Lambda (u,\,x)}
\newcommand{\lats}{\Lambda (t,\,x)}
\newcommand{\ely}{\exp (-\Lambda (t-u,\,y))}
\newcommand{\fis}{\phi (s,\,x)}
\newcommand{\oma}{\Omega ,\,{\cal A}}
\newcommand{\ikst}{{\bf x}_t}
\newcommand{\fto}{{\cal F}_t^0}
\newcommand{\ftt}{{\cal F}_t}
\newcommand{\ftn}{{\cal F}_{\infty}}
\newcommand{\ikss}{{\bf x}_s}
\newcommand{\net}{{\tilde N}_t}
\newcommand{\cpl}{{\bf C}P_{\alpha}}
\newcommand{\ptx}{P(t,\,x,\,dy)}
\newcommand{\pjt}{p^1_t}
\newcommand{\pdt}{p^2_t}
\newcommand{\pjs}{\overline{p}^1_t}
\newcommand{\pds}{\overline{p}^2_t}
\newcommand{\pat}{p^{\alpha}_t}
\newcommand{\pas}{\overline{p}^{\alpha}_t}
\newcommand{\tps}{T_t(P_x)}
\newcommand{\tpa}{T_t(P_x)_{\alpha}}
\newcommand{\gaf}{g_{\alpha_1\alpha_0}\phi (t_1,\:x)}
\newcommand{\guf}{g_{\alpha_k\alpha_{k-1}}\phi (t_k\:-\:t_{k-1},\:x_{k-1})}
\newcommand{\usa}{\hat{U}(s)\psi_0}
\newcommand{\nusa}{\|\usa\|^2}
\newcommand{\pnt}{\overline{p}_n(t)}
\title{The Piecewise Deterministic Process Associated to EEQT}
\author{Ph. Blanchard, A. Jadczyk
\footnote{permanent address: 6322 Montana Av., New Port Richey, 34653 FL}\\ and R. Olkiewicz
\footnote{permanent address: Institute of Theoretical Physics, University
of Wroc{\l}aw, PL 50-204 Wroc{\l}aw, Poland}\\
Faculty of Physics and BiBoS,
University of Bielefeld\\
Universit\"atstr. 25,
D-33615 Bielefeld\\
}
\date{ }
\maketitle
\begin{abstract}
In the framework of event enhanced quantum theory (EEQT) a probabilistic
construction of the piecewise deterministic process associated with a
dynamical semigroup is presented. The process describes sample histories
of individual systems and gives a unique algorithm generating time series
of pointer readings in real experiments.
\end{abstract}

{\bf Key words:} quantum measurements, open systems, completely positive
semigroups,  piecewise deterministic processes.
\newpage
\noindent
{\bf 1. Introduction}

One of the primary aims of quantum measurement theory is to understand the mechanism 
by which potential properties of  quantum systems become actual. This is not
an abstract or philosophical problem. Nowadays it is possible to carry
out prolonged observations of individual quantum systems. These observations
provide us with time series of data, and a complete theory must explain
the mechanism by which these time series are being generated; it must be able to "simulate" the
natural process of events generation. There are several methods of approaching this problem.
John Bell \cite{bell87a} for instance, sought a solution in hidden variable theories of Bohm
and Vigier, his own idea of beables, and also in the spontaneous localization idea of
Ghirardi, Rhimini and Weber \cite{grw85}. More recently, in a series of papers, two of us
(Ph. B. and A.J.)  \cite{blaja93a,blaja95a,blaja95b} proposed a formalism that goes in a
similar direction but avoids introducing other hidden variables beyond the wave function
itself. Our \lq\lq Event Enhanced Quantum Theory\rq\rq (in short: EEQT) describes a
consistent mode of coupling between a quantum and a classical system, in which a classical
system is one described by an Abelian algebra. 
It is an enhancement because it modifies quantum theory by adding the
new term to the Liouville equation. This allows to unify the continuous evolution
of a wave function with quantum jumps that accompany real world events. When the
coupling constant is small, events are rare, and EEQT reduces to the orthodox quantum theory.

We suggest that a  measurement process is, by
definition, a coupling of a quantum and a classical system,  where transfer of information
about quantum state to the classical recording device is mathematically modeled by a
dynamical semigroup (i.e. semigroup of completely positive and trace preserving maps) of the
total system.  It is instructive to see that such a transfer of information cannot, indeed,
be accomplished by a Hamiltonian or,  more generally, by
any automorphic evolution\footnote{For a discussion of this fact in a broader context
of algebraic theory of superselection sectors - cf. 
Landsman \cite[Sec. 4. 4]{lan}. Cf. also the no-go result by 
Ozawa \cite{oza}}.  To this end consider 
a system described by a von Neumann algebra ${\cal A}$ with
center ${\cal Z}. $ Then ${\cal Z}$ describes the classical degrees freedom of the system. 
Let $\omega $ be a state of ${\cal A}, $ and let ${\omega}|_Z$ denote its restriction to
${\cal Z}. $ Let $\alpha _t$ be an automorphic time evolution of $ {\cal A}, $ and denote
$\omega ^t=\alpha ^t (\omega ), $ where the dual evolution of states is given by $\alpha ^t
(\omega ) (A)=\omega  (\alpha _t (A)). $ Each $\alpha _t$ is an automorphism of the algebra
${\cal A} , $  and so it leaves its center invariant:  $\alpha _t: {\cal Z}\rightarrow {\cal
Z}. $ The crucial observation is that, because the evolution of states of ${\cal Z}$ is dual
to the evolution of the observables in ${\cal Z}$, and we have $\alpha^t(\omega)|_{{\cal
Z}}=\alpha^t|_{{\cal Z}}(\omega|_{{\cal Z}}),$  the restriction ${\omega ^t}|_{{\cal Z}}$
depends only on ${ \omega ^{0}}|_{{\cal Z}}.$ In other words the future state of the the
classical subsystem depends only on the past state of that subsystem and -- not on its
extension to the total system. This shows that no information transfer from the total system
to its classical subsystem is possible -- unless we use more general,  non--automorphic
evolutions. The idea of describing a quantum measurement as a two-way coupling between
quantum system and a classical system occurred before to several authors --  we mention only
the classical papers by Sudarshan \cite{su1} -- but never within the completely positive
semigroup approach.  

EEQT has several points of contact with other approaches. The mathematical model 
was a result of our studies of the papers of Jauch
\cite{jau1,jau2}, Hepp \cite{hep}, Piron \cite{pir1,pir2,pir3} , Gisin
\cite{gis1,gis2} and Araki \cite{ara}
, and also of the papers by Primas (cf. \cite{prim1,prim2}). 
It was then found that our master equation describing a coupled
quantum--classical system is of the type already well known
to statisticians. In his monographs \cite{davmha1,davmha2}
dealing with stochastic control and optimization
M. H. A. Davis, having in mind mainly queuing and insurance models,
described a special class of piecewise deterministic
processes that was later found to fit perfectly the needs of quantum 
measurement theory, and that reproduced the master equation postulated
originally by the two of us in \cite{blaja93a}. \\
In \cite{jakol95} it was shown that the special class of couplings between a
classical and quantum system leads to a unique piecewise deterministic
process with values on $E$-the pure state space of the total system.
That process consists of random jumps, accompanied by changes of a
classical state, interspersed by random periods of Schr\"odinger-type
deterministic evolution. The process, although mildly nonlinear in quantum 
wave function $\psi$, after averaging, 
recovers the original linear master equation for statistical states. 
The action of the
dynamical semigroup $T_t$ is given in terms of the process in the
following way
$$T_t(P_x)\;=\;\int P(t,\,x,\,dy)P_y,$$
where $P(t,x,dy)$ is the transition probability function of the process
and $y\to P_y$ is a tautological map, which assigns to every point
$y\in E$ a one-dimensional projector $P_y$. 
Let us discuss more precisely this connection between a dynamical 
semigroup $T_t$ and the Markov-Feller process associated with it. Suppose
${\cal M}(E)$ is a Banach space of all complex, finite, Borel measures on $E$
with norm given by $\|\mu\|\,=\,|\mu|(E)$. Let us define a map
$$\pi:\;{\cal M}(E)\to{\cal A}_{T*},\quad \pi(\mu)\;=\;\int\limits_E\mu(dx)P_x$$
where ${\cal A}_{T*}$ is the predual space of the total algebra ${\cal A}_T$. It
is clear that $\pi$ is a linear, surjective and positive map with $\|\pi\|=1$.
Therefore we can identify ${\cal A}_{T*}$ with the Banach quotient space
${\cal M}(E)/{\rm ker}\pi$. We say that a Markov process $P(t,\,x,\,dy)$ with
values in $E$ is associated with $T_t$ iff $\hat{U}_t\,=\,T_t$, where $U_t$ is
the semigroup on ${\cal M}(E)$ determined by $P(t,\,x,\,dy)$, and $\hat{U}_t$
denotes the quotient semigroup with respect to ${\rm ker}\pi$. The process is said 
to be Feller iff it preserves the space of all continuous and vanishing at infinity
functions on $E$. It means that to find the associated process we have to extend
the semigroup $T_t$ from ${\cal M}(E)/{\rm ker}\pi$ to ${\cal M}(E)$ in an invariant
way. In general, such an extension may not exists or, if it exists, may not be 
unique. However, under mild assumptions, both the existence and the uniqueness were
proved by analytical methods in ref. 22. From the point of view of the physical
interpretation the uniqueness is of great importance since it leads to a unique
description of the behavior of an individual quantum system under observation.

The main objective of this
paper is to provide a probabilistic construction of the process and
discuss some of its properties and applications. 
We present a detailed construction since the theory of piecewise
deterministic processes is not a part of the standard mathematical education.
And, we believe, it is impossible to understand the essence of EEQT without 
having even a rough idea about this theory. The paper is organized
as follows. In Section 2 the formalism for classical-quantum interactions
is presented. In Section 3 the probabilistic construction of the PD process
is described and some of its properties are analyzed. In Section 4 the
classical part of the process is discussed. We also present an example
of direct photodetection. Concluding remarks are given in Section 5.\\[4mm]
{\bf 2. The Formalism}

We start by recalling the theorem by Christensen and Evans that describes the
most general form of a generator of a completely positive semigroup of
transformations of an algebra with an non-trivial center. The theorem 
generalizes the classical
results of Gorini,  Kossakowski and Sudarshan \cite{koss} 
and of Lindblad \cite{lin} to the case of
arbitrary $C^{\star }$--algebra, and it states that
essentially the Lindblad form of the generator holds also for this more
general case. We quote the theorem for the convenience of the reader \cite{chr}: \\
{\bf Theorem 2.1.}(Christensen - Evans)
Let $\alpha_t = \exp  (L t)$ be a
norm-continuous semigroup of CP maps of a
$C^{\star }$-- algebra of operators ${\cal A}\subset {\cal B} ({\cal
H}) . $ Then there exists a CP map $\phi$
of ${\cal A}$ into the ultraweak closure ${\bar {\cal A}}$
and an operator $K\in {\bar {\cal A}}$ such that the
generator $L$ is of the form: 
$$L (A) = \phi  (A) + K^{\star }A + AK $$ 

Let us apply this theorem to the case of ${\cal A}$ being a von Neumann
algebra,  and the maps $\alpha_t$ being normal.  Then $\phi$ can be
also taken normal.  We also have ${\bar {{\cal A}}} = {\cal A} , $
so that $K\in {\cal A} . $ Let us assume that $\alpha_t  (I) = I $ or, 
equivalently,  that $L (I)=0 . $  It is convenient to introduce $
H=i (K-K^{\star })/2 \in {\cal A}, $ then from $L (I)=0$ we get $K+K^{\star
}=-\phi  (I) , $ and so $K=-iH-\phi  (I)/2 . $ Therefore we have 
$$L (A) = i\left[H, A\right]+\phi  (A) -\{ \phi  (I) , A\}/2$$          
where $\{\,  ,  \,  \}$ denotes anticommutator. 

We now apply the above formalism to the hybrid system which is a direct
product of the classical and quantum mechanical one. The physical idea
behind such a model is that a quantum measurement is to be defined as a particular
coupling between a quantum and a classical system. We continuously observe the
classical system, notice changes of its pure states (we call these changes "events")
and from these we deduce properties of the coupled quantum system. 
Details can be found in refs. 4 and 5. One can think
of events as `clicks' of a particle counter, sudden changes of the pointer velocity,
changing readings on an apparatus LCD display. The concept of an event is
of course an idealization - like all concepts in a physical theory. Let us
consider the simplest situation corresponding to a finite set of possible
events. The space of pure states of our classical system $C$, denoted by ${\cal S}_c$, has
$m$ states, labeled by $\alpha=1,\ldots,m$. Statistical states of $C$ are probability
measures on ${\cal S}_c$ -- in our case just sequences $p_\alpha\geq 0, \sum_\alpha
p_\alpha=1$. 

The algebra of observables of $C$ is the algebra $\ca_c$ of complex
functions on ${\cal S}_c$ -- in our case just sequences $f_\alpha, \alpha =1,\ldots,m$ of
complex numbers. We use Hilbert space language even for the description of the classical
system. Thus we introduce an $m$-dimensional Hilbert space $\calh_c$ with a fixed basis, and
we realize $\ca_c$ as the algebra of diagonal matrices $F=\dia(f_1,\ldots,f_m)$. Statistical
states of $C$ are then diagonal density matrices $\dia(p_1,\ldots,p_m)$, and pure states of
$C$ are vectors of the fixed basis of $\calh_c$. Events are ordered pairs of pure states
$\alpha\rightarrow\beta$, $\alpha\neq\beta$. Each event can thus be represented by an
$m\times m$ matrix with $1$ at the $(\alpha,\beta)$ entry, zero otherwise. There are $m^2-m$
possible events. Let us point out that important here is the discreteness of the classical
system not its finiteness. We can easily generalize the above to the case when the classical
points form, for example, the set of natural numbers. Then the classical algebra becomes
$l^{\infty}$ (uniformly bounded sequences) while statistical states are positive elements
from $l^1$ (summable sequences).\\ We now come to the quantum system. Let $Q$ be the quantum
system whose bounded observables are from the algebra $\ca_q$ of bounded operators on a
Hilbert space $\calh_q$. In this paper we will assume  $\calh_q$ to be {\em finite
dimensional}\ . Pure states of $Q$ are unit vectors in $\calh_q$; proportional vectors
describe the same quantum state. They form a complex projective space ${\bf C}P(\calh_q)$
over $\calh_q$. Statistical states of $Q$ are given by non--negative density matrices
${\hat\rho}$,  with $\ttr ({\hat\rho})=1$.

Let us now consider the total system $T=Q\times C$.
For the algebra $\ca_T$ of observables of $T$
we take the tensor product of algebras of observables of $Q$ and $C$:
$\ca_T =\ca_q\otimes\ca_c$. It acts on the tensor product
$\calh_q\otimes\calh_c=\oplus_{\alpha=1}^m\calh_\alpha$, where
$\calh_\alpha\approx\calh_q.$ Thus
$\ca_T$ can be thought of as algebra of {\em diagonal} $m\times m$
matrices $A=(a_{\alpha\beta})$, whose entries are quantum operators:
$a_{\alpha\alpha}\in \ca_q$, $a_{\alpha\beta}=0$ for $\alpha\neq\beta$.
Statistical states of $Q\times C$ are given by $m\times m$ diagonal matrices
$\rho=\dia(\rho_1,\ldots,\rho_m)$ whose entries are positive operators
on $\calh_q$, with the normalization $\ttr (\rho)=\sum_\alpha\ttr
(\ra)=1$.
Duality between observables and states is provided
by the expectation value $<A>_\rho=\sum_\alpha \ttr (\aaa\ra)$.

We will now generalize slightly our framework. Indeed, there is no need
for the quantum Hilbert spaces $\calh_\alpha$, corresponding to different
states of the classical system, to coincide. We will allow them to
be different in the rest of this paper. Intuitively such a generalization
corresponds to the idea that a phase transition can accompany the event. We denote
$n_\alpha=dim(\calh_{\alpha})$.

We consider now dynamics. It is normal in quantum theory that classical
parameters enter quantum Hamiltonian. Thus we assume that
quantum dynamics, when no information is
transferred from $Q$ to $C$, is described by Hamiltonians
$\ha :\calh_\alpha \longrightarrow \calh_\alpha$,
that may depend on the actual state of $C$ (as indicated by the index
$\alpha$).
We will use matrix notation and write $H=\dia(\ha)$.
Now take the classical system. It is discrete here.
Thus it can not have continuous time dynamics of its own.

The coupling of $Q$ to $C$ is specified by a
matrix $V=(\gab)$, where $\gab$ are linear operators:
$\gab :\calh_\beta \longrightarrow \calh_\alpha$.
We assume  $g_{\alpha\alpha}=0$. This condition expresses the simple
fact: we do not need dissipation without receiving information i.e without an event. 
To transfer information from $Q$ to $C$ we need a non--Hamiltonian term which provides
a completely positive (CP) coupling. As in \cite{blaja95a,blaja95b} we consider
couplings for which the evolution
equation for observables and for states is given by the Lindblad  form:
$${\dot A}_\alpha=i[\ha,\aaa]+\sum_\beta \gba^\star
A_\beta \gba - {1\over2}\{\la,\aaa\},\label{eq:lioua}$$
or equivalently:
$${\dot \rho}_\alpha=-i[\ha,\ra]+\sum_\beta \gab
\rho_\beta \gab^\star - {1\over2}\{\la,\ra\},\label{eq:liour}$$
where
$$\la=\sum_\beta \gba^\star \gba$$
The above equations describe statistical behavior of ensembles. Individual
sample histories are described by the following algorithm:\\
{\em Suppose that at time $t_0$ the system is described by a normalized
quantum state vector $\psi_0$ and a classical state $\alpha$. Then choose
a uniform random number $p\in [0,\,1]$, and proceed with the continuous
time evolution by solving the modified Schr\"odinger equation
$$\dot{\psi}_t\:=\:(-iH_{\alpha}\:-\:\frac{1}{2}\la )\psi_t$$
with the initial wave function $\psi_0$ until $t\,=\,t_1$, where $t_1$ is
determined by
$$\int\limits_{t_0}^{t_1} (\psi_t,\,\la\psi_t)dt\:=\:p$$
Then jump. When jumping, change $\alpha\to\beta$ with probability
$$p_{\alpha\to\beta}\:=\:\|\gba\psi_{t_1}\|^2/(\psi_{t_1},\,\la\psi_{t_1})$$
and change
$$\psi_{t_1}\to\psi_1\:=\:\gba\psi_{t_1}/\|\gba\psi_{t_1}\|.$$
Repeat the steps replacing $t_0,\,\psi_0,\,\alpha$ with $t_1,\,\psi_1$
and $\beta$.}\\
This leads to a stochastic process, in which the randomness appears as point
events i.e. there is a sequence of random occurrences at random times 
$T_1<T_2<\,...$, but there is no additional component of uncertainty between
these times. It consists of a mixture of deterministic motion and random
jumps. 
A class of such processes is called piecewise deterministic processes
(PDP) \cite{davmh3}. The motion between jumps is determined by a complete vector
field $X$ on the pure state space $E$ of the total system. The jump
mechanism is determined by two further components: a jump rate $\lambda$
and a transition kernel $Q$. The vector field $X$ generates a flow $\phi(t,
\,x)$ in $E$, which is given by $\phi(t,\,x)\:=\:\gamma_x(t)$, where 
$\gamma_x(t)$ is the integral curve of $X$ starting at point $x\in E$. The
jump rate is a measurable function 
$\lambda:E\to {\bf R}_+$ such that
for any $x\in E$ the mapping $t\to\lambda\circ\phi(t,\,x)$ is integrable
at least near $t\:=\:0$. The set of those $x\in E$ for which $\lambda(x)\:=
\:0$ we denote by $E_0$.
The transition kernel $Q:{\cal B}(E)\times E\to
[0,\,1]$ satisfies the following conditions:\\
a) $Q(E,\,x)\:=\:1\quad\forall x\in E$,\\
b) $Q(\{x\},\,x)\:=\:0$ if $x\in E\setminus E_0$ and $Q(\{x\},\,x)\:=\:1$
for $x\in E_0$,\\
c) $\forall\Gamma\in{\cal B}(E)$ the map $x\to Q(\Gamma,\,x)$ is 
measurable.\\
Here ${\cal B}(E)$ denotes the Borel $\sigma$-algebra on $E$. In our case
$E\:=\:\dot{\bigcup}{\bf C}P_{\alpha},\:\alpha\,=\,1,\,2,...,m$ and
we have the following formulas for $X,\:\lambda$ and $Q$:\\
$$Xf(\psi,\,\alpha)\:=\:\frac{d}{dt}f(\frac{\exp(-iH_{\alpha}\:-\:\frac{1}{2}
\la)\psi}{\|\exp(-i\ha\:-\:\frac{1}{2}\la)\psi\|},\,\alpha)|_{t=0}$$
$$\lambda(\psi,\,\alpha)\:=\:<\psi,\,\la\psi>$$
$$Q(d\phi,\,\beta;\,\psi,\,\alpha)\:=\:\frac{\|\gba\psi\|^2}{\lambda(\psi,\,
\alpha)}\delta(\phi\:-\:\frac{\gba\psi}{\|\gba\psi\|})d\phi$$
if $(\psi,\,\alpha)\in E\setminus E_0$
and $\delta(\phi)d\phi$ denotes the Dirac measure.\\
The triple $(X,\,\lambda,\,Q)$ is called local characteristic of the process.
Its infinitesimal generator is given by
$${\cal L}f(x)\:=\:Xf(x)\:+\:\lambda(x)\int\limits_E [f(y)\:-\:f(x)]Q(dy,\,
x)$$ and produces sample paths exactly such as described by the above
algorithm. It is worth noting that there are no correlations between jump times $T_i$,
$i\in{\bf N}$. However, as we will see in the next section, the survival function of
random variable $T_1$ is exponential, given by
$$F_t\;=\;\exp(-\int\limits_0^t\lambda(\ikss )ds)$$
where $\ikss$ denotes the piecewise deterministic process.\\[4mm]
{\bf 3. The PD process}

In this section we present the detailed construction of the process introduced
in Section 2 and investigate some of its properties. General references on
stochastic processes are \cite{protter,jacod}. Probabilistic concepts can be
found in \cite{neveu,grimm}.

At first we construct a probabilistic space $(\oma )$ (compare \cite{yushk}
for a similar construction for Markov decision processes).
Let $\Omega$ be a set
of all sequences $(t_0,x_0;t_1,x_1;\ldots )$, which are finite or infinite,
and such that $t_0=0$, $t_n\leq t_{n+1}$, $t_n\in\dot{\bf R}_+=\:[0,\infty ]$,
$x_n\in E$ for all $n\in {\bf N}\cup\{0\}$. If a sequence is finite i.e.
$\omega =\:(t_0,x_0;\ldots ,t_n,x_n)$ then we put
$$t_{n+1}=t_{n+2}=\ldots =\infty,\quad\quad x_{n+1}=x_{n+2}=\ldots =x_n$$
It follows that $\Omega$ can be embedded into an infinite product space
$\prod\limits_{n=0}^{\infty} \Omega_n$, where $\Omega_0=\{0\}\times E$ and
$\Omega_n=\dot{\bf R}_+\times E$. On each $\Omega_n$ we have a natural 
$\sigma$-algebra ${\cal A}_n$ given by ${\cal B}(\dot{\bf R}_+)\otimes\boe$. 
We define a $\sigma$-algebra ${\cal A}$ on $\Omega$ as $(\otimes_
{n=0}^{\infty} {\cal A}_n)\vert_{\Omega}$.\\ 
Now let us construct a family of
probabilistic measures $P_x$ on $(\oma )$ with respect to an initial state
$x\in E$. They will be determined by the deterministic drift $\phi$, the
jump rate $\lambda$ and the transition kernel $Q$. Because we want to use
the Ionescu Tulcea theorem \cite{neveu} we have to define transition kernels
between $(\Omega_n,\;{\cal A}_n)$ and $(\Omega_{n+1},\;{\cal A}_{n+1})$.
We do it step by step. \\ On $\Omega_0$ we take the
Dirac measure $P_0=\delta_x$. Let $\lats :=\int^t_0 \lambda (\fis)ds$ and
let us define
$$F_x(t_1)\;=\;1-\exp (-\Lambda (t_1,\,x))$$
$$K_x(t_1,\,dx_1)\;=\;Q(dx_1,\,\phi (t_1,\,x))$$
As the transition kernel between $(\Omega_0,{\cal A}_0)$ and $(\Omega_1,
{\cal A}_1)$ we take
$$P_0^1(x,\,B_1\times\Gamma_1)\;=\;\int\limits_{B_1}\int\limits_{\Gamma_1}
K_x(t_1,\,dx_1)dF_x(t_1)$$
for any $B_1\in {\cal B}(\dot{\bf R}_+)$ and any $\Gamma_1\in\boe$. 
In the second step we define
$$F_{(t_1,x_1)}(t_2)\;=\;\cases{0,&if $t_1>t_2$\cr 1-\exp 
(-\Lambda (t_2-t_1,\,x_1)),&if $t_1\leq t_2$}$$
$$K_{(t_1,x_1)}(t_2,\,dx_2)\;=\;Q(dx_2,\,\phi (t_2-t_1,\,x_1))$$
and put
$$P_2^1(t_1,\,x_1;\,B_2\times\Gamma_2)\;=\;\int\limits_{B_2}\int\limits_{
\Gamma_2} K_{(t_1,x_1)}(t_2,\,dx_2)dF_{(t_1,x_1)}(t_2)$$
It is clear that $P_2^1$ is a transition kernel between $(\Omega_1,{\cal A}_1
)$ and $(\Omega_2,{\cal A}_2)$. In the similar way we construct higher
kernels $P^n_{n+1}$. By Ionescu Tulcea theorem 
there is a unique probabilistic measure $P_x$ on $(\prod\limits_{n=0}^{
\infty} \Omega_n,\;\otimes_{n=0}^{\infty} {\cal A}_n)$ such that for every
measurable rectangle
$A\:=\:A_0\times A_1\times\ldots\times A_n\times \Omega_{n+1}
\times\ldots$ the following identity 
$$P_x[A]\;=\;\delta_x(A_0)\int\limits_{A_1} P_0^1 (x;\,dt_1,\,dx_1)\cdots\int
\limits_{A_n} P^{n-1}_{n}(t_{n-1},\,x_{n-1};\,dt_n,\,dx_n)$$
is satisfied. It is clear from the above formula that $P_x$ is concentrated on
$\Omega_x\:=\:\{\omega\in\Omega :\;x_0=\:x\}$, $x\in E$.
Moreover $P_x$ is measurable
with respect to $x$. \\ To investigate properties of the above measure
let us define a sequence of measurable random variables
$$T_n:\Omega_x\to\dot{\bf R}_+\quad T_n(\omega)\;=\;t_n,\quad X_n:\Omega_x\to E
\quad X_n(\omega)\;=\;x_n$$
The distributions of $T_0$ and $X_0$ are Dirac measures concentrated in
$\{0\}$ and $\{x\}$ respectively. 
The distribution $dF_{T_1}$ of $T_1$ is given by
$$P_x[T_1\leq t]\;=\;1-\exp (-\lats )$$
and the conditional expectation of $X_1$ given $T_1$ equals to
$$E_x[1_{\{X_1\subset\Gamma\}}\vert T_1]\;=\;Q(\Gamma,\,\phi (T_1,\,x))$$
Here $1_{\{\cdot\}}$ denotes an indicator function of a given set. The above
equation can be also written as
$$dF_{X_1|T_1}(y\vert t)\;=\;Q(dy,\,\fit ),$$
where the left hand side is the conditional distribution of $X_1$. For
arbitrary $n\in {\bf N}$ we have the following formulas:
$$E_x[1_{\{T_{n+1}\leq t\}}\vert T_n,\,X_n]\;=\;\cases{0&if $t<T_n$\cr
1-\exp (-\Lambda (t-T_n,\,X_n))&if $t\geq T_n$}$$
$$E_x[1_{\{X_{n+1}\subset\Gamma\}}\vert X_n,\,T_{n+1}]\;=\;Q(\Gamma,\,
\phi (T_{n+1},\,X_n))$$
It follows that $P_x[T_1=0]\:=\:0$ so $T_1>0$ a.s. Because, after a jump,
process starts again so $T_n<T_{n+1}$ a.s. for every $n$. 
This fact can be also derived from the following equality:
$$P_x[T_{n+1}-T_n>s]\;=\;E_x[\exp (-\Lambda (s,\,X_n))]$$
It means that a set of paths with two or more simultaneous jumps has
zero probability. Moreover,
because $Q(\{x\},\,x)\:=\:0$ for every $x\in E\setminus E_0$
so with probability one the process can not
jump to the state it is deterministically approaching.
There are no jumps from the set $E_0$ at all.\\
Let us calculate some physically interesting probabilities. For example the
probability that there is no jump up to time $t$ equals to
$$P_x[T_1>t]\;=\;\exp (-\int\limits_0^t \lambda (\fis )ds)$$
Because $$P_x[T_2>t]\;=\;P_x[T_1>t]\:+\:E_x[1_{\{T_1\leq t\}}\exp (-\Lambda
(t-T_1,\,X_1)]$$
and, on the other hand,
$$P_x[T_2>t]\;=\;P_x[T_2>t\wedge T_1\leq t]\;+\;P_x[T_2>t\wedge T_1>t]$$
so the probability that exactly one jump happens up to time $t$ is given by
$$P_x[T_2>t\wedge T_1\leq t]\;=\;\int\limits_0^t 1_{\{u\leq t\}}dF_{T_1}(u)
\int_E \ely dF_{X_1|T_1}(y\vert u)\;=$$
$$\int_0^t\int_E \lambda (\fuks )\exp (-\luks )\ely Q(dy,\,\fuks )du$$
Now let us define a random variable $T_{\infty}=\lim_{n\to\infty} T_n$. For
every $t<T_{\infty}$ we construct the process $\ikst$ by putting
$$\ikst (\omega)\;=\;\phi(t-T_k(\omega),\,X_k(\omega))\quad if \quad
T_k(\omega)\leq t<T_{k+1}(\omega)$$
In general we can have the process with the lifetime. We show that in our
case, due to the boundedness of the jumping rate, 
$T_{\infty}=\infty$ a.s. Let $C=\sup_{x\in E} \lambda(x)$. Then
for every $t>0$
$$\sup\limits_{x\in E} (1-\exp (-\lats )\leq 1\:-\:e^{-Ct}$$
Let us fix $t$ and denote $C_1\:=\:1\:-\:e^{-Ct}$, which is strictly less
than 1. Then
$$P_x[T_{n+1}\leq t]\;=\;E[1_{\{T_n\leq t\}}(1-\exp (-\Lambda(t-T_n,\,X_n))]
\leq C_1P_x[T_n\leq t]\leq C_1^{n+1}$$
by induction. It implies that
$$P_x[\bigcap_{n=0}^{\infty}\{T_n\leq t\}]\;=\;\lim\limits_{n\to\infty}
P_x[T_n\leq t]\;=\;0$$
It follows that $\ikst$ is defined for all $t\in {\bf R}_+$ and 
is a {\it cadlag} process i.e. possesses
right continuous with left limits paths. \\
To end the construction of ingredients needed for a Markov process
let us introduce a natural filtration on $\Omega_x$ given by
$\fto =\sigma\{\ikss ,\;s\leq t\}$ and take ${\cal F}_{\infty}^0=\vee_t
\fto$. Let $\ftt$ and ${\cal F}_{\infty}$ denote the $P_x$-completion of
$\fto$ and ${\cal F}_{\infty}^0$ respectively. Because, after a jump, the
process evolves deterministically, so the filtration $(\ftt )_{0\leq t\leq
\infty}$ is right continuous. Hence the filtered
probability space $(\Omega,\,\ftn,\, P_x,\,\ftt )$
satisfies the usual hypothesis for every $x\in E$ and $\ikst$ is an adapted 
and {\it cadlag} process. Because the distribution of $T_1$ depends only on the current
state $\ikst$ and, after a jump, process starts again so
$(\Omega,\,\ftn ,\,P_x,\,\ftt ,\,\ikst )$ is a strong Markov process with
infinite lifetime.

Next we show another important property of the process $\ikst$, namely the
quasi-left-continuity. 
Let us define a random set $\triangle =\:\{(t,\omega):\;{\bf x}_{t^-}
\neq\ikst \}$, where ${\bf x}_{t^-}$ is the left limit of $\ikst$. Then
$$\nu(\omega;\,dt,\,dx)\;=\;\sum\limits_s 1_{\triangle}(s,\,\omega)\delta_{
(s,\ikss (\omega))}(dt,\;dx),$$
where $\delta_{(s,x)}$ is the Dirac measure on ${\bf R}_+\times E$ 
concentrated in $(s,\;x)$, is an integer-valued random measure. It leads to
a simple point process $\net$ given by
$$\net\;=\;\nu([0,\,t]\times E)\;=\;\sum\limits_{n=1}^{\infty} 1_{\{T_n\leq
t\}}$$
Because $T_{\infty}=\:\infty$ a.s. so $\net$ is a.s. finite valued. It is
also integrable because
$$E_x[\net ]\;=\;\sum\limits_{n=1}^{\infty} nP_x[T_n\leq t]\leq\;\sum\limits_
{n=1}^{\infty} nC_1^n\;<\;\infty$$
Moreover it was shown in \cite{davmh3} that the compensator of $\net$ is equal
to $\int_0^t \lambda(\ikss )ds$ and so $M_t:=\:\net -\int_0^t \lambda (\ikss
)ds$ is an $(P_x,\,\ftt )$-martingale. Using this fact it can be calculated
that the dual predictable projection of $\nu$ is given by
$$\nu^p(\omega ;dt,\,dx)\;=\;Q(dx,\,\ikst (\omega))\lambda(\ikst (\omega))dt$$
Thus $\nu^p(\omega;\,\{t\},\,E)\:=\:0$ and so $\ikst$ is quasi-left-continuous
\cite{jacod}. Thus we proved that $\ikst$ is a Hunt process. Moreover $\ikst$ is
a Feller process i.e. the transition kernel of $\ikst$ generates a strongly
continuous semigroup of contractions on the space of all continuous functions
on $E$, see \cite{gatarek1,gatarek2}. \\[4mm]
{\bf 4. Stochastic representation of the classical system}

In this section we discuss some properties of the stochastic process
associated with the measuring apparatus. Let $C$ be a state space of the
classical system i.e. $C\:=\{1,\,2,\,...,m\}$. Let us define a
$\{0,\,1\}$-valued process $\pat$ by
$$\pat (\omega)\;=\;\delta^{\alpha}_{\pi (\ikst (\omega))},$$
where $\delta_{\beta}^{\alpha}$ is the Kronecker delta and $\pi$ denotes
the canonical projection $\pi :\;E\to C$. By $P(t,\,x,\,\Gamma )$, $x\in E$
and $\Gamma\subset E$ we denote the transition kernel of the process $\ikst$.
It was shown in \cite{jakol95} that $P(t,\,x,\,\Gamma )$ is associated with the
dynamical semigroup $T_t$ and so
$$ T_t(P_x)\;=\;\int\limits_E P_y\ptx$$
Here $P_y$ is the one-dimensional projector corresponding to $y\in E$ i.e.
$P_y=\:|y><y|$. We show that the average of $\pat$ gives the probability of
finding the total system at time $t$
in a classical state $\alpha$. Let $\pas\:=\:E[\pat ]$.
Then $$\pas\;=\;\int\limits_E \delta^{\alpha}_{\pi (y)}\ptx\;=\:\int\limits_
{\cpl}{\rm Tr}(P_y)\ptx\;=$$
$${\rm Tr}(\int\limits_{\cpl} P_y\ptx )\;=\;{\rm Tr}(\tpa )$$
Now we derive a differential equation for $\pas$. Let us start with the
following example.\\
{\bf Example 1.} Let $C\:=\{1,\,2\}$. Then a change of the process $\pat$,
$\alpha =1,\;2$, is given by
$$d\pjt\;=\;-p^1_{t^-}d\net\;+\;p^2_{t^-}d\net ,\quad p_0^1\;=\;1$$
$$d\pdt\;=\;-p^2_{t^-}d\net\;+\;p^1_{t^-}d\net ,\quad p^2_0\;=\;0,$$
where $\net$ is the counting process introduced in the previous section.
Solving the above equations we get
$$\pjt\;=\;\frac{1+(-1)^{\net}}{2},\quad \pdt\;=\;\frac{1-(-1)^{\net}}{2}$$
Because $M_t\:=\:\net -\int_0^t \lambda(\ikss )ds$ is a martingale so we
get the following equations for averages $\pas$:
$$d\pjs\;=\;E[(-\pjt +\pdt )\lambda(\ikst )]dt$$
$$d\pds\;=\;E[(-\pdt +\pjt )\lambda(\ikst )]dt$$
When the intensity is a constant function equal to $\lambda$ they reduce to
$$d\pjs\;=\;\lambda (-\pjs +\pds )dt,\quad d\pds\;=\;\lambda (-\pds +\pjs )dt$$
with solutions given by
$$\pjs\;=\;\frac{1+e^{-2\lambda t}}{2},\quad\pds\;=\;\frac{1-e^{-2\lambda t}}
{2}$$
{\bf Proposition 4.1.}
$$d\pas\;=\;-E[\lambda_{\alpha}(\ikst )]dt\;+\;\sum\limits_{\beta\neq\alpha}
E[\|g_{\alpha\beta}\ikst\|^2]dt$$
Proof: Because $\pas\:=\:{\rm Tr}(\rho_{\alpha})$, $\rho_{\alpha}\:=\:\tpa$ so
$$\frac{d\pas}{dt}\;=\;{\rm Tr}(\dot{\rho}_{\alpha})\;=\;{\rm Tr}(-i[H_{\alpha},\,\rho
_{\alpha}]\;-\;\frac{1}{2}\{\Lambda_{\alpha},\,\rho_{\alpha}\}\;+$$
$$\sum\limits_{\beta\neq\alpha} g_{\alpha\beta}\rho_{\beta}g^{\ast}_{\alpha
\beta})\;=\;{\rm Tr}(\Lambda_{\alpha})\rho_{\alpha})\;+\;\sum\limits_{\beta\neq\alpha}
{\rm Tr}(g^{\ast}_{\alpha\beta}g_{\alpha\beta}\rho_{\beta})$$
On the other hand
$$E[\lambda_{\alpha}(\ikst )]\;=\;\int\limits_{\cpl} <y|\Lambda_{\alpha}|y>\ptx
\;=\;{\rm Tr}(\Lambda_{\alpha}\tpa )$$
$$E[\|g_{\alpha\beta}\ikst\|^2]\;=\;\int\limits_{{\bf C}P_{\beta}} <y|g^{\ast}
_{\alpha\beta}g_{\alpha\beta}|y>\ptx\;=\;{\rm Tr}(g^{\ast}_{\alpha\beta}g_{\alpha
\beta}T_t(P_x)_{\beta})$$
so the assertion follows. $\Box$\\
The advantage of this stochastic representation of ${\rm Tr}(\tpa )$ is that we can
predict the future of the classical system if we know its past. Let us point
out that the classical component of $\ikst$ usually is not a Markov process.

Let us assume that we start at $t=0$ with a quantum state $x\in {\bf C}P_{
\alpha 0}$, and up to the present we have observed the following classical
trajectory 
$$(t_0=0,\,\alpha_0),\,(t_1,\,\alpha_1),...,(t_k\leq\,t,\,\alpha_k)$$
Then the probability $p_{\alpha}$ that the next jump will go to $\alpha$ can
be obtained as follows. Let us calculate
$$x_1\;=\;\frac{\gaf}{\|\gaf\|},\quad ...\quad x_k\;=\;\frac{\guf}{\|\guf\|}$$
and ${\bf x}_t\:=\:\phi (t-t_k,\,x_k)$ for $t\geq t_k$. Then
$$p_{\alpha}\;=\;E[\|g_{\alpha\alpha_k}({\bf x}_{T_{k+1}})\|^2]\;=\;
\int\limits_0^{\infty} dF_{T_{k+1}|T_k,X_k}(t|t_k,\,x_k)\|g_{\alpha\alpha_k}
(\ikst )\|^2\;=$$
$$\int\limits_{t_k}^{\infty} dt\exp (-\int\limits_{t_k}^t \lambda (\ikss )ds)
\|g_{\alpha\alpha_k}(\ikst )\|^2$$
These probabilities can be also
used to determine an initial quantum state. Let us
assume that we start at $t=0$ with a classical index $\alpha_0$ and with one
of the following pure quantum states $x^i\in {\bf C}P_{\alpha 0}$, $i=1,\;2,
...,n$. The probability that the first jump will change the classical index
onto $\alpha$ is given by
$$p^i_{\alpha}\;=\;E[\|g_{\alpha\alpha_0}({\bf x}_{T_1})\|^2],\quad  T_0\;=\;0,
\quad X_0\;=\;x^i$$
In the similar way we calculate probabilities $p^i_{\alpha 2\alpha 1}$ that
the first jump will go to $\alpha_1$ and the second one to $\alpha_2$ and so
on. Taking appropriate $g_{\beta\alpha}$ we can make these probabilities
significantly different for each initial quantum state $x^i$ and thus conclude
which one is the most probable by observing the classical trajectory.\\
{\bf Example 2}. Let us consider the fluorescent photons emitted by a
single, two-level atom that is coherently driven by an external 
electromagnetic field. It is known that the quantum system evolves from
the ground state in a dissipative way. When a photoelectric count is
recorded by a photoelectric detector (we assume the detector efficiency
to be equal to one), the atom returns to the ground state with the emission of
one photon. Thus, after the emission of each photon, the atom starts its
evolution from the same state. We describe this situation using the
probabilistic framework introduced in the previous sections.\\
The quantum system as a two-state system is represented by $2\times 2$
complex matrices. The classical system, which counts emitted photons we
describe by an infinite sequence of numbers $n\,=\,0,\,1,\,2,...$ Hence
the state space of the total system is equal to
$$E\;=\;\bigcup\limits_{n=0}^{\infty}{\bf C}P^2$$
The time evolution of the quantum system is described (for every classical
index $n$) by the modified Schr\"odinger equation
$$\dot{\psi}_t\;=\;-i\hat{H}\psi_t\;=\;(-iH\:-\:\frac{1}{2}\Lambda)\psi_t,$$
where $\Lambda\,=\,\gamma A^*A$ and
$$A\;=\;\left(\begin{array}{cc}0&0\\1&0\end{array}\right)$$
The coupling operators $g_{nm}$ are given by $g_{n+1,n}\,=\,A$ and $g_{mn}\,=\,0$ if
$m\neq n+1$.
A solution for $\psi_t$ can be written as $\psi_t\,=\,\hat{U}(t)\psi_0$,
where \cite{breuer}
$$\hat{U}(t)\;=\;e^{-it\hat{H}}\;=\;e^{-\gamma t/4}\left(\begin{array}{cc}
\cos\mu t\:-\:\frac{\gamma}{4\mu}\sin\mu t&i\frac{\Omega}{2\mu}\sin\mu t\\
i\frac{\Omega}{2\mu}\sin\mu t&\cos\mu t\:+\:\frac{\gamma}{4\mu}\sin\mu t
\end{array}\right)$$
Here $\gamma$ is the relaxation rate, $\Omega$ is Rabi frequency and
$\mu\,=\,\sqrt{\Omega^2\:-\:(\gamma/2)^2}$. The ground state $\psi_0$
is given by $$\psi_0\;=\;\left(\begin{array}{c}0\\1\end{array}\right)$$
The deterministic flow is defined by
$$\phi(t,\,(\psi_0,\,n))\;=\;\frac{\hat{U}(t)\psi_0}{\|\hat{U}(t)\psi_0\|}$$
for every $n\in{\bf N}\cup\{0\}$. The jump rate $\lambda$ is given by
$\lambda((\psi,\,n))\;=\;<\psi,\,\Lambda\psi>$ for all $n$ and the
transition kernel
$$Q(d\phi,\,m;\,\psi,\,n)\;=\;\delta^m_{n+1}\delta(\phi\:-\:\psi_0)d\phi$$
Because of the uniqueness of a jump after the emission of a photon the
classical component of the piecewise deterministic process of the total
system is also a Markov process. Let us derive the distribution of the
waiting time between jumps. In this case it is exactly the distribution
of the random variable $T_1$. Thus
$$F(t)\;=\;1\;-\;\exp[-\Lambda(t,\,(\psi_0,\,0))\;
=\;1\;-\;\exp[-\int\limits_0^t\lambda(\phi(s,\,(\psi_0,\,0))ds]$$
$$=\;1\;-\;\exp[-\int\limits_0^t<\frac{\usa}{\|\usa\|},\,\gamma A^*A\frac{\usa}{
\|\usa\|}>ds]$$ Because
$$\frac{d}{ds}\nusa\;=\;<-i\hat{H}\usa,\,\usa>\:+\:<\usa,\,-i\hat{H}\usa>$$
$$=\;<\usa,\,i(\hat{H}^*\:-\:\hat{H})\usa>\;=\;-<\usa,\,\gamma A^*A\usa>$$
so we obtain that
$$F(t)\;=\;1\;-\;\exp[\int\limits_0^t(\frac{d}{ds}\log\nusa)ds]\;=\;
1\;-\;\nusa$$
Its density is given by
$$f(t)\;=\;\gamma\|A\hat{U}(t)\psi_0\|^2\;=\;\gamma\frac{\Omega}{4\mu^2}
\sin^2(\mu t)\exp(-\gamma t/2)$$
It is exactly the waiting time density obtained in \cite{carmi}.\\
Now let us consider the time evolution of the averages of the classical
components of the process $\pnt$. In the present context they have a
simple interpretation: $\pnt$ is the probability that $n$ photoelectric
counts are recorded in the time interval $[0,\,t]$. Hence $\pnt\,=\,
P[T_n\leq t\wedge T_{n+1}>t]$ and so
$$\overline{p}_0(t)\;=\;\|\hat{U}(t)\psi_0\|^2$$
$$\overline{p}_1(t)\;=\;\int\limits_0^t\overline{p}_0(s)f(t-s)ds\;=\;
(\overline{p}_0\ast f)(t)$$
$$\overline{p}_2(t)\;=\;\int\limits_0^t ds_2\overline{p}_0(s_2)\int
\limits_0^{t-s_2}f(s_1)f(t-s_2-s_1)ds_1\;=\;(\overline{p}_0\ast f\ast f)(t)$$
and so on. In the above we extended functions $\overline{p}_0(t)$ and
$f(t)$ to the whole real line $(-\infty,\,\infty)$ by putting value zero
for negative arguments. The sign $\ast$ denotes the convolution. Taking
the Laplace transform of $\pnt$ with respect to variable $t$ we obtain that
$$\hat{p}_n(\lambda)\;=\;\int\limits_0^{\infty}e^{-\lambda t}\pnt dt\;=\;
\hat{p}_0(\lambda)[\hat{f}(\lambda)]^n$$
which coincides with the formula given in \cite{carmi} (see also \cite{breuer}).

The above simple example demonstrates how the piecewise deterministic dynamics
works in practice. It also shows that EEQT reproduces known results in certain
domains. But it also predicts more in other domains, where the orthodox quantum
theory is silent. For example the EEQT algorithm has been applied to tunneling
time problems \cite{Palao}. In particular, traversal and reflection times of
electrons through a one-dimensional barrier have been calculated in \cite{Rusch}. 
There EEQT gives predictions which can be tested and
compared with those stemming from other approaches. In this respect, the orthodox
quantum theory gives no prediction at all. It is also possible using the framework
of EEQT to consider a simultaneous measurement of noncommuting observables \cite{Jadcarib}.
In the simplest case of a simultaneous measurement of several spin projections for
a spin 1/2 quantum particle, the piecewise deterministic process turned out to be
a non-linear version of the Barnsley's iterated function system and led to the
chaotic behavior and fractal structure on the space of pure states \cite{Jastrz}.\\[4mm]
{\bf 5. Concluding remarks}

The crucial concept underlining our approach to quantum measurement is that of a
classical and irreversible event. This is taken into account by including
from the beginning classical degrees of freedom. From the structural point
of view such a coupling (EEQT) consists of the following essential ingredients:\\
- tensoring of a non-commutative quantum algebra of observables with a
classical commutative algebra (or, more generally, taking the classical
Abelian algebra as the center of the total algebra of observables), \\
-replacing Schr\"odinger unitary dynamics by a completely positive
semigroup describing the time evolution of ensembles,\\
- interpreting the continuous time evolution of statistical states in terms
of a piecewise deterministic process with values in the pure state space
of the total system,\\
- applying the uniqueness theorem for deducing the piecewise deterministic
algorithm generating sample path of an individual system.\\
This (EEQT) gives a minimal extension of the quantum theory which
ensures the flow of information from the quantum system to the classical
variables. Further, EEQT provides a way to calculate numbers needed in real
experiments and also allows for natural mathematical modeling of feedback
during experiments with quantum systems.

This coupling with classical variables does not mean we are taking a step 
backward into classical mechanics. We are only saying that not all is quantum
and there are elements of nature that are not, and clearly cannot be, described by a quantum wave
function. Even if this is viewed as an assumption, it is firmly confirmed by 
experiments which show that we are living in the world of facts, not in the world of 
possibilities. Thus, for this aspect, which is clearly not reducible to quantum degrees of 
freedom, we have adopted the term "classical variables." However, this does not imply
that we intend to impose any strict restrictions on their nature; they may prove to 
be related, for example, to gravity or even to consciousness. If there is no 
interaction between these classical variables and quantum degrees of freedom, both 
systems may prove to evolve separately according to their own equation of motion. 
It may be that when they interact, a new dissipative operation appears which results in the 
irreversibility of the evolution and leads to collapses in measurement situations. 
It is worth noting that there is a fundamental difference between the classical 
variables in EEQT and additional parameters in hidden variable theories.
Hidden variable theories
use microscopic variables that are hidden indeed from our observations. EEQT 
deals with classical variables that can be observed. In fact, it states that
there are no other variables that can be directly observed. They are a direct 
counterpart of physics on the other side of the Heisenberg-von Neumann cut. 
Further, in hidden variable theories there is no back action of these variables 
on the wave function. In EEQT there is such an action.

Let us also discuss two possible generalizations of the above framework. The first
one concerns the dimension of the quantum Hilbert space. Here, for the sake of
simplicity, we have used only finite dimensional Hilbert spaces, but from the 
construction it is quite clear that this assumption is not essential. We can admit 
infinite dimensional Hilbert spaces, but this requires additional work so that the 
formulas are well defined.  For example, the infinite series of operators must be 
convergent. Further, we can allow the Hamiltonian operator to be unbounded. Also the 
existence of the deterministic flow can be established since ${\bf C}P(\calh_q)$ is an
infinite dimensional Hilbert manifold; that is, it can be covered by a family
of open sets each of which is homeomorphic to an open ball in a Hilbert space.

The second generalization is connected to the discreteness of the classical system.
Although the basic applications of EEQT concerns measurement processes, it can
also encompass a non-trivial interaction between a quantum system and a
classical continuous one. The most transparent example is the SQUID-tank model,
which consists of an electric oscillatory circuit coupled
via a mutual inductance to a superconducting ring.
In that system the oscillatory circuit acts as an external
flux source for the SQUID ring, which induces a screening current in the
ring. This screening current is coupled back to the classical circuit due
to the mutual inductance. It results in the modification of the differential
equation for the damped classical harmonic oscillator by the expectation value
of the superconducting screening current operator. Now the classical phase space is a symplectic
manifold ${\bf R}^2$ and the equation of motion is given by 
$$C\ddot{\phi}\;+\;\frac{1}{R}\dot{\phi}\;+\;\frac{1}{L}\phi\;=\;I(t)$$
where $\phi$ is the classical flux variable. The quantum object evolves according to Hamiltonian
$$H\;=\;\frac{\hat{Q}^2}{2C}\;+\;\frac{(\hat{\Phi}-\Phi_x)^2}{2\Lambda}\;-\;\hbar\nu\cos(2\pi\frac{
\hat{\Phi}}{\Phi_0})$$
where $\hat{Q}\,=\,-i\frac{d}{d\Phi}$ is the momentum operator, $\hat{\Phi}$ is the position operator,
$\Phi_0\,=\,\frac{h}{2e}$ and $\Phi_x$ is the external magnetic flux. Suppose $\rho_t$ is an evolving
statistical operator of the total system and define the so-called collective classical variable
$$\overline{x}_t\;=\;\int x(\mbox{Tr}\rho_t(x,\,p))dxdp$$
Then its evolution equation takes form (we omit the physical constants):
$$\ddot{\overline{x}}_t\;+\;\dot{\overline{x}}_t\;+\;(1\,+\,\|f\|_{L^2}^2)\overline{x}_t\;=\;
I(t)\;+\;\|f\|_{L^2}^2<\hat{I}_s>_{\hat{\rho}}$$
where $\hat{I}_s$ is the screening current operator, $f$ is a function monitoring the strength of the
coupling and $\hat{\rho}$ is the reduced density matrix obtained by tracing over classical variables.
Here $<\hat{I}_s>_{\hat{\rho}}$ denotes the expectation value of $\hat{I}_s$ in state $\hat{\rho}$.
For more details see ref. 40. It is worth emphasizing that the above modification of the evolution
of the collective classical variable, which was postulated by experimental physicists (see eq.
(12) in ref. 41), has been derived in a rigorous mathematical way within the framework of EEQT.
Finally, using EEQT, a possible influence of the quantum matter on the classical gravitational
field has been demonstrated. It was achieved by the change of a dynamical path of the classical
particle moving freely along a geodesic curve when interacting with the quantum system.  In average
the classical evolution equation is perturbed by the 
expectation value of the quantum position operator \cite{Olk99}. The probabilistic description of 
the dynamics in this case was presented in ref. 43 and the back action of the classical variables
on the quantum system, resulting in its non-trivial asymptotic behavior, was discussed in ref 44.

To sum up: EEQT, although it is a phenomenological model, is very promising and can be successfully
applied to a large class of physical phenomena.\\[4mm] 
{\bf Acknowledgments} We would like to thank the referee for his remarks which helped us to improve
the clarity of the paper. Two of the authors (A.J. and R.O.) would like to thank A.
von Humboldt Foundation for the financial support. 

\end{document}